\documentclass[twocolumn]{revtex4}
\usepackage{graphicx}
\usepackage{epstopdf}
\usepackage{amsmath}
\usepackage{appendix}
\usepackage{color}
\begin{document}

\title{MSene: A new large family of two-dimensional transition metal sulfide with MXene structure}
	
\author{Shu-Xiang Qiao$^{1}$, Yu-Lin Han$^{1}$, Na Jiao$^{1}$, Meng-Meng Zheng$^{1}$, Hong-Yan Lu$^{1,\footnote{E-mail: hylu@qfnu.edu.cn}}$, and Ping Zhang$^{1,2}$}
\affiliation{$^1$School of Physics and Physical Engineering, Qufu Normal University, Qufu 273165, China \\$^2$Institute of Applied Physics and Computational Mathematics, Beijing 100088, China}

\begin{abstract}  	
In this work, we theoretically report a new large family of two-dimensional (2D) transition metal sulfides $M$$_{2}$S with MXene structure in 2H and 1T phases, which we name as MSene. Twenty-four out of fifty-eight MSenes are proved to be stable. Notably, this family includes twelve superconducting (SC) materials, seven SC topological metals (SCTMs), four charge density wave (CDW) materials, and five magnetic materials including one ferromagnetic (FM) and four antiferromagnetic (AFM) materials. For example, 2H-Mo$_{2}$S is a SCTM which exhibits SC critical temperature ($T_{c}$) of 10.2 K and nontrivial topological properties; 1T-Hf$_{2}$S is a CDW material with the CDW originating from electron-phonon coupling. The CDW can be suppressed by compressive strain, leading to the emergence of superconductivity; 2H-Cr$_{2}$S and 1T-Mn$_{2}$S show FM and AFM properties, respectively. Thus, the new large family we predicted shows rich physical properties and significantly expands the repertoire of 2D materials. It serves as a novel platform for investigating the competition or coexistence of multiple orders such as SC, CDW, FM, AFM and topological orders in 2D materials.
\end{abstract} 

\maketitle

\textit{Introduction.} Since the successful fabrication of graphene in 2004 \cite{graphene-experiment}, 2D systems have garnered significant attention due to their distinctive properties, which differ from corresponding three-dimensional structures. In recent years, various 2D materials have been investigated, such as hydrides \cite{jacs,cuh2,h-mgb2,h-c,h-b}, carbides \cite{c1,c2,graphene-sc,T-graphene,biphenylene1,biphenylene2,lic6,alc8}, borides \cite{b1,b2,b3}, nitrides \cite{n1,bn,bn3,h-bn}, transition metal dischalcogenides (TMDs) \cite{tmd1,tmd2,tmd3,tmd4,tmd5}, and MXenes \cite{mxene1,mxene2,mxene3,mxene4}. These materials typically exhibit outstanding electronic, mechanical, catalytic, adsorbing and optical properties, making them a focal point of research in physics, chemistry, and materials science.

TMDs are a widely studied class of 2D materials, whose chemical formula can be written as $M$$X$$_{2}$ ($M$ = transition metal, $X$ = chalcogenide element). TMDs possess unique physical properties, including intrinsic magnetism, CDW, and superconductivity, making them highly attractive for applications in high-performance electronic devices, catalysis, and metal electrode contacts \cite{tmd1,tmd4}. Experimental methods for preparing TMDs are both simple and diverse \cite{tmd6,tmd7}. For instance, MoS$_{2}$ has been successfully fabricated through hydrothermal and thermal decomposition synthesis methods \cite{mos2-experiment}, and has lead to wide-ranging applications \cite{mos2app1,mos2app2,mos2app3}.

\begin{figure*}\centering\label{fig:structure}\includegraphics[width=15cm]{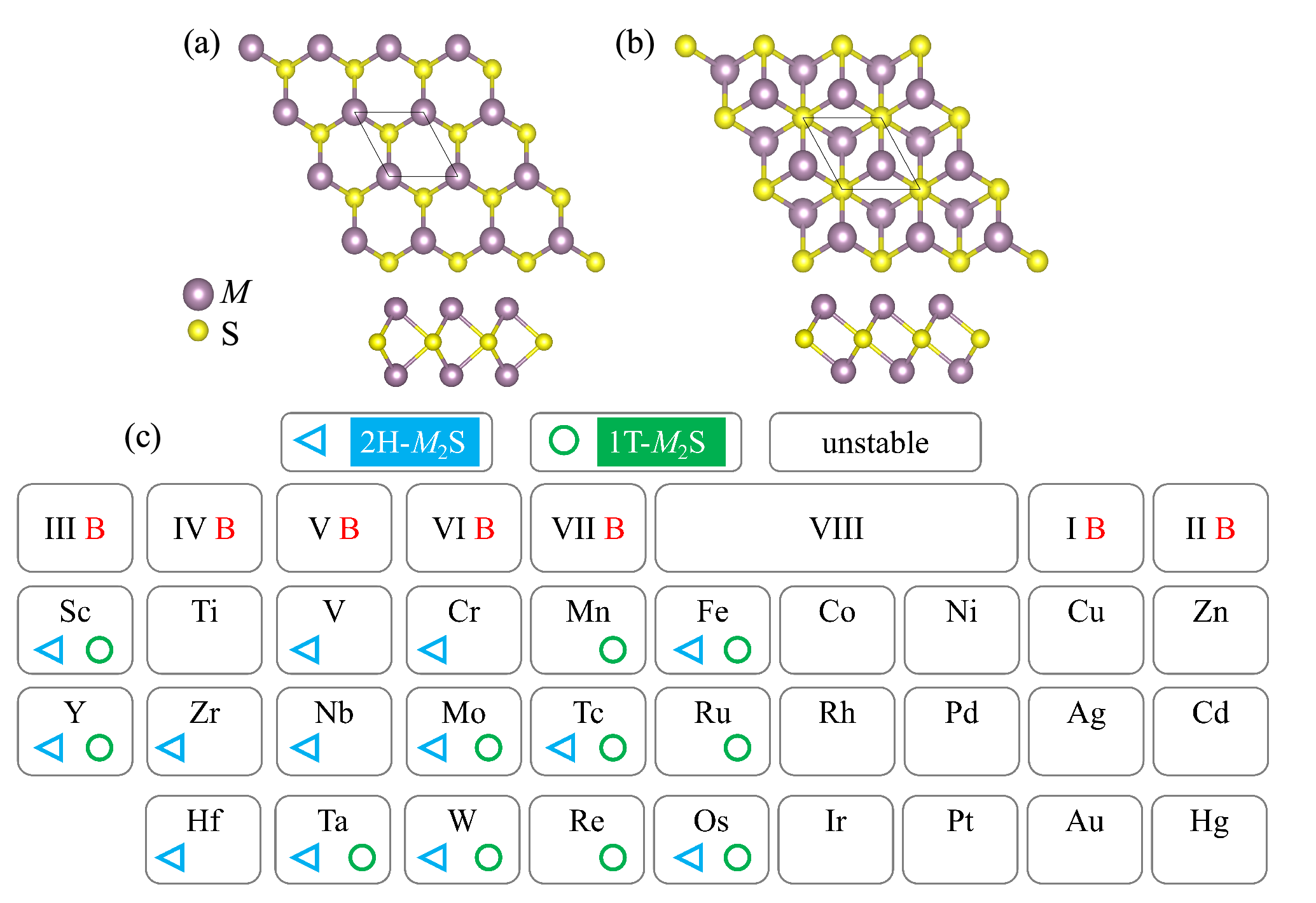}
	\caption{Top and side views of (a) 2H, and (b) 1T-MSenes. The primitive cell is shown by the black solid line. (c) Distribution of stable 2D MSenes $M$$_{2}$S where $M$ represents transition metal. For clarity, blue triangles and green circles represent stable MSenes in 2H and 1T phase, respectively.}\end{figure*}

Apart from TMDs, MXene represents another class of extensively studied 2D materials \cite{mxene5,mxene6,mxene7,mxene8}. MXene can be obtained by etching A layer from MAX phase. The simplest structure in MXene is $M$$_{2}$$X$ ($M$ = transition metal, $X$ = C or N). With its high metal conductivity and excellent chemical stability \cite{mxene5}, diverse physical properties of MXene have been experimentally and theoretically investigated, including superconductivity \cite{mo2c,yxz,wanghao}, band topology \cite{topology}, and magnetism \cite{mag1,mag2}, etc.

Considering the diverse and simple preparation methods of TMDs and good conductivity of MXenes, this work theoretically reports a new large family of transition metal sulfides $M$$_{2}$S ($M$ = transition metal) combining the element composition of TMDs and the stoichiometry of MXenes, and we name it MSene. In the MSene family, we considered the $M$$_{2}$S in 2H and 1T phases and found twenty-four out of fifty-eight materials to  be stable. Among them, there are twelve SC materials, seven SCTMs, four CDW materials, and five magnetic materials with one in FM and four in AFM ground state. The new large family we predicted greatly enriches the family of 2D materials, and provides a new platform for studying the competition or coexistence of several orders in 2D materials.

\begin{figure*}\centering\label{fig:summary}\includegraphics[width=18cm]{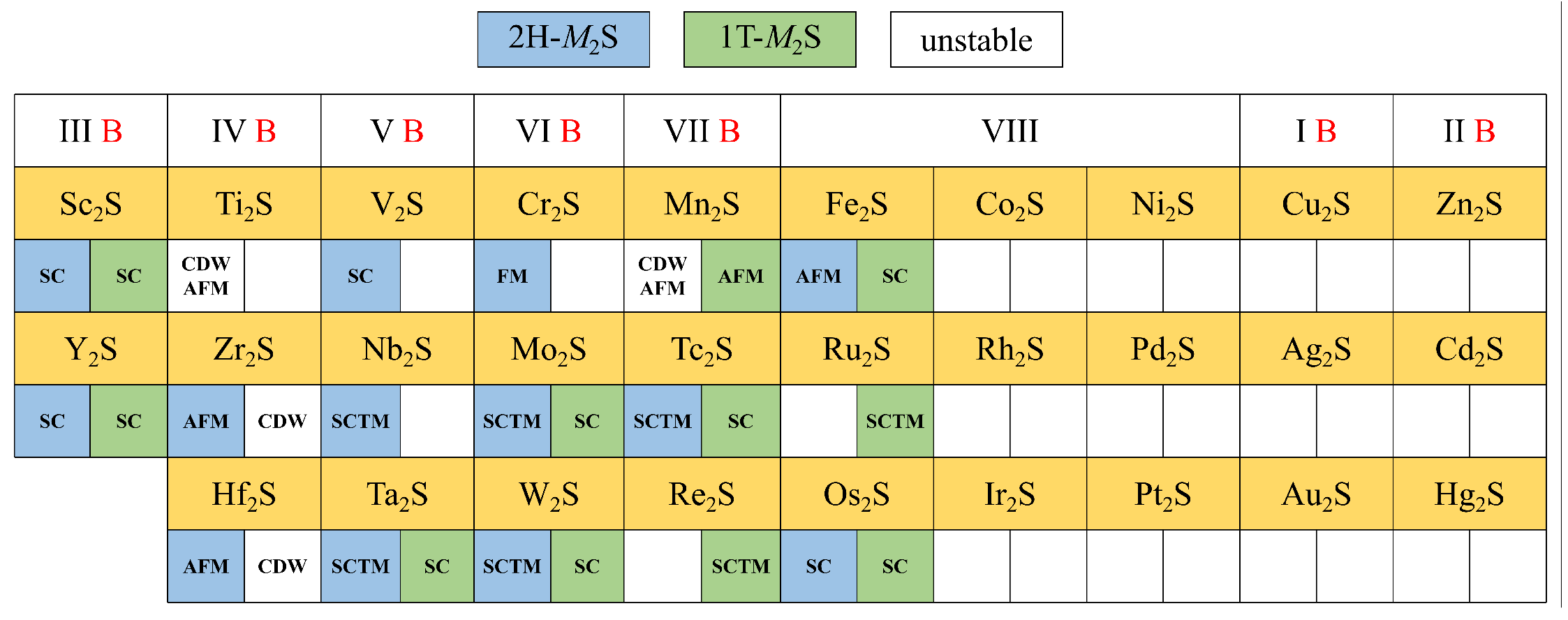}
	\caption{ A summary of properties of MSenes. The blue, green and white blocks represent the stable 2H phase, stable 1T phase and unstable structures, respectively.}\end{figure*}

\textit{Structures and stability.} Figures 1(a) and 1(b) show the top and side views of 2H and 1T-MSenes $M$$_{2}$S. The space group of 2H- and 1T-MSenes are $P$-$6m2$ and $P$-$3m1$, respectively. Unlike TMDs such as MoS$_{2}$, MSenes consist of two layers of transition metal atoms and one layer of S atoms sandwiched between the $M$ layers. The computational methods and details are shown in the Supplemental Material (SM) \cite{sm}. The relaxed lattice parameters of stable MSenes are listed in Table S1 in the SM \cite{sm}. The charge density difference in Fig. S1 \cite{sm} reveals that $M$ atoms lose electrons while S atoms gain electrons. The electron localization function (ELF) in Fig. S1 \cite{sm} indicates that electrons are distributed around the S atoms, suggesting the formation of ionic bonds.

Then, we demonstrate the dynamic and thermodynamic stability of MSenes using phonon spectra, cohesive energy $E_{coh}$, and $ab$-$initio$ molecular dynamics (AIMD) simulations. Figure 1(c) presents a complete diagram of stability information. The absence of imaginary frequency in their phonon spectra indicates the dynamic stability, with twenty-four out of fifty-eight 2D MSenes materials identified as dynamically stable ones. The phonon spectra of all MSenes are shown in Fig. S2 \cite{sm}.

\begin{figure*}\centering\includegraphics[width=18cm,height=10.5cm]{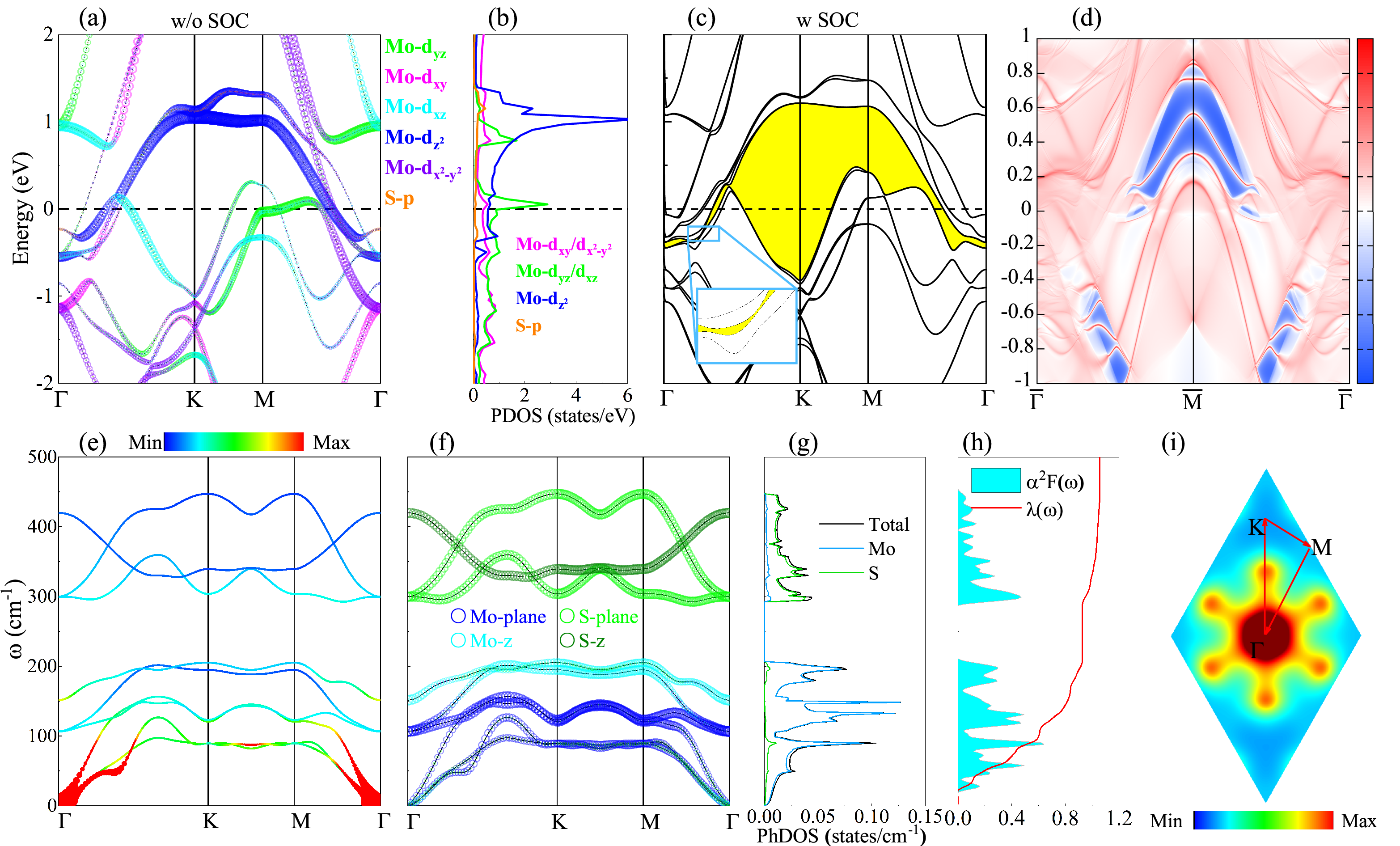}
	\caption{ Band topology and SC properties of 2H-Mo$_{2}$S. (a) Orbital-projected electronic band structure without SOC along high-symmetry line $\Gamma$-$K$-$M$-$\Gamma$. (b) PDOS without SOC. (c) Band with SOC. (d) Edge states. (e) Phonon dispersion weighted by the magnitude of $\lambda_{\textbf{q}\nu}$ (EPC for phonon mode ${\textbf{q}\nu}$). (f) Phonon dispersion weighted by the vibration modes of each atom. (g) Phonon DOS. (h) Eliashberg spectral function $\alpha^{2}F (\omega)$ and EPC $\lambda (\omega)$. (i) The integrated EPC distributions in the plane $q_{z}$ = 0.}\end{figure*}

The $E_{coh}$ of dynamically stable MSenes are illustrated in a bar graph in Fig. S3 and Table S2 \cite{sm}. Notably, 2H-Hf$_{2}$S in the MSene family has been successfully fabricated in experiment \cite{hf2s-exp}, and the $E_{coh}$ of most MSenes are lower than 2H-Hf$_{2}$S, as shown in Fig. S3 and Table S2 \cite{sm}. Furthermore, the calculated $E_{coh}$ of MSenes are lower than some experimentally fabricated 2D materials, such as Cu$_{2}$Si ($-$3.46 eV/atom) \cite{cu2si1,cu2si2}, GaSe ($-$2.81 eV/atom) \cite{gas}, and GaS ($-$3.62 eV/atom) \cite{gas}. AIMD simulations reveal that the dynamically stable MSenes also exhibit thermodynamic stability at room temperature or above, as demonstrated in Fig. S4 \cite{sm}. In a word, twenty-four MSenes shown in Fig. 1(c) exhibit good stability, providing promising avenues for future experimental exploration.

Based on the stable MSenes, we discuss their interesting and exciting properties. By analyzing the electronic bands and DOS, it is found that they have bands crossing the Fermi level and exhibit metallicity, as shown in Fig. S5 \cite{sm}. An overview of the physical properties of the stable MSenes is presented in Fig. 2. In the following, we select 2H-Mo$_{2}$S, 1T-Hf$_{2}$S, 2H-Cr$_{2}$S, and 1T-Mn$_{2}$S as representatives of SCTM, CDW, FM, and AFM materials for detailed analysis, respectively.

\begin{figure*}\centering\includegraphics[width=18cm,height=9cm]{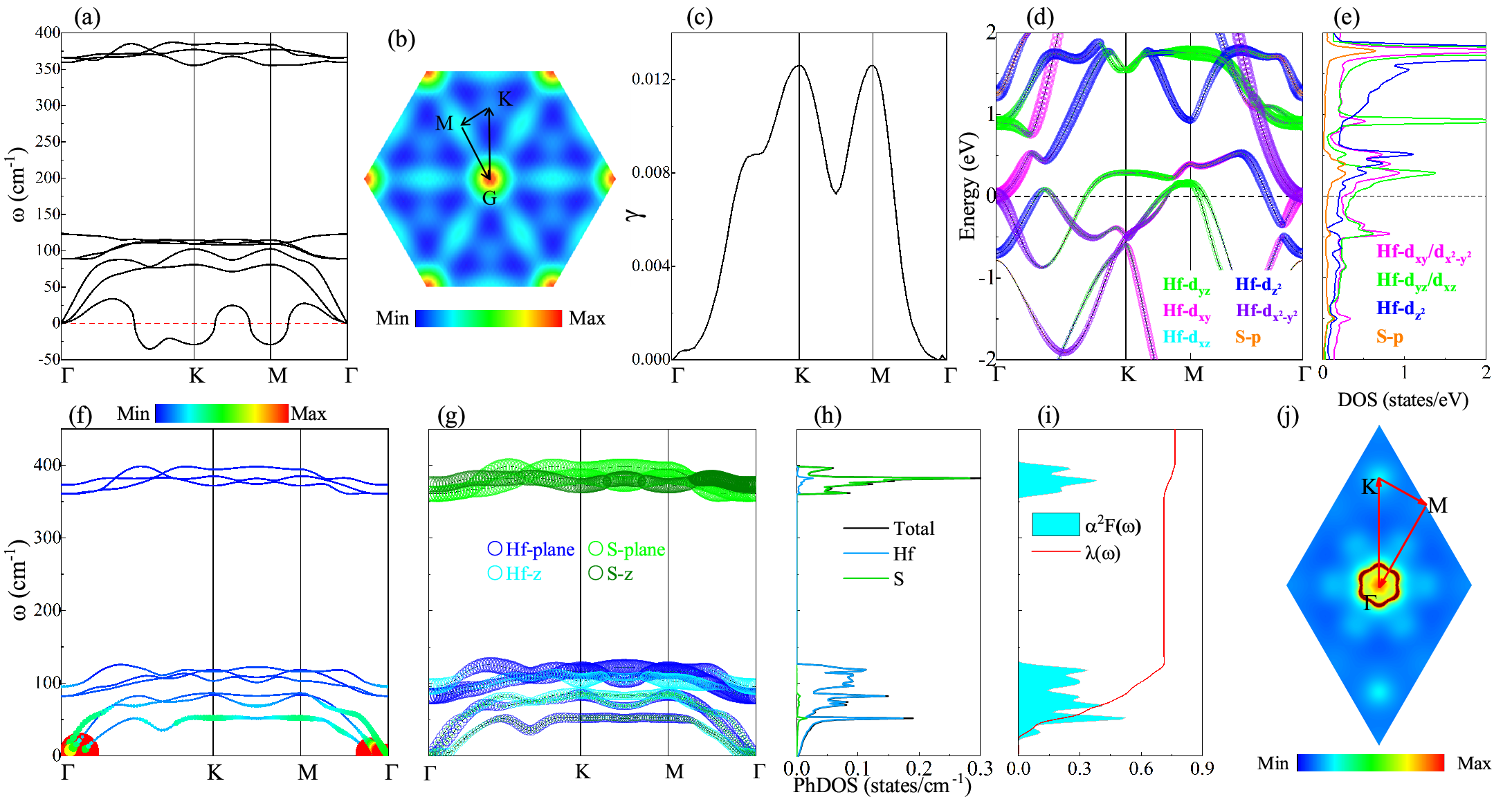}
	\caption{Properties of pristine 1T-Hf$_{2}$S (a-c) and 4$\%$ compressive strained 1T-Hf$_{2}$S (d-j). (a) Phonon spectrum. (b) Nesting function. (c) Phonon linewidth of the lowest acoustial phonon branch. (d) Orbital-projected electronic band structure. (e) PDOS. (f) Phonon dispersion weighted by the magnitude of $\lambda_{\textbf{q}\nu}$ (EPC for phonon mode ${\textbf{q}\nu}$). (g) Phonon dispersion weighted by the vibration modes of each atom. (h) Phonon DOS. (i) Eliashberg spectral function $\alpha^{2}F(\omega)$ and EPC $\lambda(\omega)$. (j) The integrated EPC distributions in the plane $q_{z}$ = 0.}\end{figure*}

\textit{Band topology and superconductivity.} The orbital-projected electronic band without spin-orbit coupling (SOC), and the partial density of states (PDOS) of 2H-Mo$_{2}$S are shown in Figs. 3(a) and 3(b). It can be observed from Fig. 3(a) that the bands near the Fermi level are mainly contributed by $d_{yz}$ and $d_{xz}$ orbitals. As shown in Fig. 3(b), the contribution of Mo-$3d$ orbitals near the Fermi level is much greater than that of S atoms, thus the metallicity of 2H-Mo$_{2}$S is mainly contributed by the Mo-$d_{yz}$/$d_{xz}$ orbitals. As shown in Fig. 3(c), when SOC is considered, the band near the Fermi level opens a continuous gap, which is highlighted in yellow. The presence of continuous band gap throughout the entire Brillouin zone enables the definition of topological invariants $Z_{2}$. Then, the calculation of the $Z_{2}$ yields a value of 1, proving that 2H-Mo$_{2}$S is topologically nontrivial. Additionally, the presence of edge states (red trace in Fig. 3(d)) also proves its nontrivial topology. The bands without and with SOC, and edge states of the other six SCTMs are presented in Fig. S6 \cite{sm}.

The superconductivity of 2H-Mo$_{2}$S is investigated with the results shown in Figs. 3(e)-3(i). In Figs. 3(f) and 3(g), phonon DOS (PhDOS) is divided into two distinct regions, with the low-frequency region mainly contributed by the vibration of Mo atoms, while the vibrational modes of S atoms are mainly in the high-frequency region, owing to the larger mass of Mo atoms. As shown in Fig. 3(e), the main contribution of its electron-phonon coupling (EPC) is from acoustic branches around $\Gamma$, consistent with Fig. 3(i), and primarily from in-plane vibrations of Mo atoms, as depicted in Fig. 3(f). The strong EPC in the low-frequency region ($w$ $<$ 120 cm$^{-1}$) contributes 62.4 $\%$ to the EPC constant $\lambda$ as shown in Fig. 3(h). The calculated EPC constant $\lambda$ and $T_{c}$ of 2H-Mo$_{2}$S are 1.06 and 10.2 K, respectivity. The superconductivity mainly originates from the coupling between Mo-$d_{yz}$/$d_{xz}$ electrons and the low-frequency in-plane vibration of Mo atoms. Considering that 2H-Mo$_{2}$S also show nontrivial topological band structure, it is a potential topological superconductor, which maybe used in fault-tolerant quantum computation.

The phonon spectra weighted by the magnitude of $\lambda_{\textbf{q}\nu}$ and Eliashberg spectral function of other SC MSenes are presented in Fig. S7 \cite{sm}. The calculated SC information are summarized in Table S3 \cite{sm}, including logarithmically averaged phonon frequency $\omega_{log}$, total EPC constant $\lambda$, SC critical temperature $T_{c}$, sommerfeld constant $\gamma$, specific heat jump at $T_{c}$ ${\Delta C\left(T_c\right)}$, SC gap at 0 K ${\Delta(0)}$, critical magnetic field at 0 K $H_c(0)$, isotope coefficient $\alpha$, and $\frac{2 \Delta(0)}{k_{{B}}T_{{c}}}$.  It is noteworthy that the calculated $\frac{2\Delta(0)}{k_{{B}}T_{{c}}}$ and $\alpha$ of most MSenes are very close to the values given by conventional BCS theory ($\frac{2\Delta(0)}{k_{{B}} T_{{c}}}$ $\approx$ 3.53 and $\alpha$ $\approx$ 0.5), confirming that MSenes are conventional phonon-mediated superconductors. Moreover, the $T_{c}$ of several MSenes are higher than that of most TMDs and MXenes, as shown in Table S4. The calculated SC parameters provide reliable data for comparison with future experiments.

\textit{CDW property.} CDW refers to a periodic fluctuation in charge density that occurs in crystals. It is observed that there are large imaginary frequencies near $K$ and $M$ in the phonon spectrum of 1T-Hf$_{2}$S, as depicted in Fig. 4(a), which is a typical phenomenon in CDW materials. CDW primarily originates from two sources: Fermi surface (FS) nesting and EPC. The imaginary part of the electronic susceptibility and the phonon linewidth $\gamma$ of the lowest acoustic phonon mode are calculated to reveal the source of CDW. Figure 4(b) shows the FS nesting. If the source of CDW is FS nesting, the nesting function exhibits relative maximum at the positions imaginary frequency appears in the phonon spectrum. However, in the case of 1T-Hf$_{2}$S, there is no relative maximum near $K$ and $M$. Consequently, it can be inferred that the CDW of 1T-Hf$_{2}$S does not originate from FS nesting. Then, the phonon linewidth $\gamma$ of the lowest phonon mode is calculated, which directly reflects the EPC intensity of this branch, as illustrated in Fig. 4(c). It is evident that two peaks are located at $K$ and $M$, which aligns with the position of the imaginary frequency shown in Fig. 4(a). This correlation indicates that EPC is the origin of CDW in 1T-Hf$_{2}$S.

Then, the effect of biaxial compressive strain on CDW is investigated. Biaxial compressive strain is defined as $\varepsilon$ = ($a$ $-$ $a_{0}$)/$a_{0}$ $\times$ 100\%, where $a_{0}$ and $a$ represent the lattice constants. The phonon spectrum of 4$\%$ compressive strained 1T-Hf$_{2}$S is depicted in Fig. 4(f). It is apparent that biaxial compressive strain can suppress CDW. The corresponding electronic states near the Fermi level are mainly composed of Hf-$d_{yz}$/$d_{xz}$ orbitals and the contribution of S is much smaller, presented in Figs. 4(d) and 4(e). Due to the competitive relationship between CDW and superconductivity, the superconductivity of 4$\%$ compressive strained 1T-Hf$_{2}$S is further investigated.

As shown in Figs. 4(f) and 4(i), the EPC around $\Gamma$ is the strongest, consistent with Fig. 4(j). The strong EPC in the low-frequency region ($w$ $<$ 110 cm$^{-1}$) contributes approximately 85.2$\%$ to the total EPC constant $\lambda$. From Figs. 4(d), 4(e), 4(g) and 4(h), it can be seen that the superconductivity of 1T-Hf$_{2}$S mainly originates from the coupling between Hf-$d_{yz}$/$d_{xz}$ electrons and the low-frequency in-plane orbital vibrations of Hf. The EPC constant $\lambda$ and $T_{c}$ of 1T-Hf$_{2}$S are 0.77 and 4.5 K, respectivity. 1T-Zr$_{2}$S shows similar phenomenon to 1T-Hf$_{2}$S, and its CDW also originates from EPC. Under 2$\%$ compressive strain, the CDW is suppressed and exhibits superconductivity with $T_{c}$ of 2.1 K in Fig. S8 \cite{sm}. Due to the magnetism of 2H-Ti$_{2}$S and 2H-Mn$_{2}$S, we do not investigate the complex relationship between magnetism, CDW, and superconductivity in this work, and will study it later.

\begin{figure}\centering\label{fig:mag}\includegraphics[width=9cm,height=6cm]{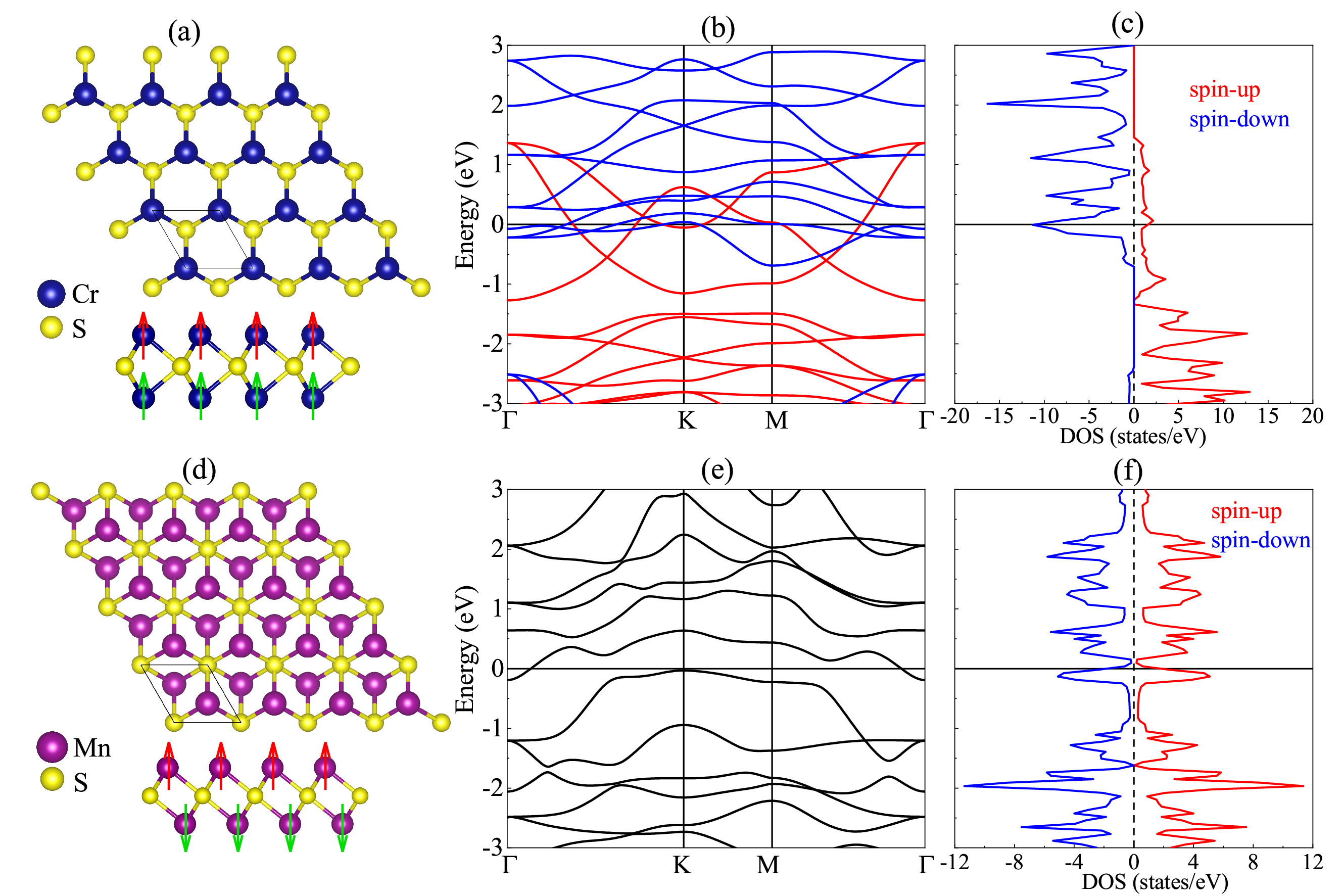}
	\caption{ (a) Top and side views, (b) Band structure, (c) DOS of 2H-Cr$_{2}$S. (d) to (f) are similar to (a) to (c), but for 1T-Mn$_{2}$S. The arrows in (a) and (d) indicate the the direction of magnetic moment.}
\end{figure}

\textit{Magnetic property.} Magnetic materials are crucial for various applications. FM materials have important applications in magnetic memory and other fields due to their external magnetic moment, and AFM materials can be used as spin electron materials in the field of information storage. Within the MSenes family, 2H-Hf$_{2}$S has been experimentally prepared and theoretically predicted to be an AFM material with surface anomalous Hall effect and electric field-induced layer Hall effect \cite{hf2s-exp, hf2s-cal}. The results for 2H-Hf$_{2}$S in this study are consistent with previous research \cite{hf2s-cal}. Additionally, 2H-Cr$_{2}$S exhibits FM properties, with the magnetic moment of the Cr atoms on both layers pointing in the same direction with the value of 3.3 $\mu$B, as shown in Fig. 5(a). Figs. 5(b) and 5(c) illustrate the band and DOS of 2H-Cr$_{2}$S, with red indicating spin-up and blue indicating spin-down. Additionally, 1T-Mn$_{2}$S displays AFM properties, with the magnetic moments of the Mn atoms on both layers being opposite in direction with the value of 3.7 $\mu$B, as shown in Fig. 5(d). Due to the AFM nature, the bands and DOS for spin-up and spin-down overlap with each other in Figs. 5(e) and 5(f). The bands of other magnetic materials are presented in Fig. S4, and their magnetic moments are shown in Table S5 \cite{sm}. These results provide theoretical data reference for future experimental research.

\textit{Conclusion.} In conclusion, we have predicted twenty-four stable MSenes with diverse physical properties, including band topology, superconductivity, CDW, and magnetism. Among them, there are twelve SC materials, seven SCTMs, four CDW materials, and five magnetic materials, including one FM and four AFM materials. Notably, the 2H-Mo$_{2}$S among SCTMs exhibits a $T_{c}$ of 10.2 K, surpassing that of most TMDs and MXenes. In CDW materials, there is a competition between CDW and superconductivity. The CDW originating from EPC can be suppressed by compressive strain and exhibit superconductivity. Therefore, our results significantly expands the repertoire of 2D materials and provides a new platform for studying the competition or coexistence of several orders such as SC, CDW, FM, AFM and topological orders in 2D materials. The predicted MSenes may provide rich applications in condensed matter physics, material physics, and device physics in the future.

\textit{Acknowledgements.} This work is supported by the National Natural Science Foundation of China (Grant Nos. 12074213, and 11574108), the Major Basic Program of Natural Science Foundation of Shandong Province (Grant No.ZR2021ZD01), the Natural Science Foundation of Shandong Provincial (Grant No. ZR2023MA082), and the Project of Introduction and Cultivation for Young Innovative Talentsin Colleges and Universities of Shandong Province.


\begin{thebibliography}{99}
		
		
\bibitem{graphene-experiment} K. S. Novoselov, A. K. Geim, S. V. Morozov, D. Jiang, Y. Zhang, S. V. Dubonos, I. V. Grigorieva and A. A. Firsov, Science {\bf 306}, 666–669 (2004).
\bibitem{jacs} X. C. Zhou, Y. Hang, L. Liu, Z. H. Zhang, and W. L. Guo, J. Am. Chem. Soc. \textbf{141}, 7899 (2019).
\bibitem{cuh2} X. Yan, S. C. Ding, X. H. Zhang, A. Bergara, Y. Liu, Y. C. Wang, X. F. Zhou, and G. C. Yang, Phys. Rev. B {\bf 106}, 014514 (2022).
\bibitem{h-mgb2} J. Bekaert, M. Petrov, A. Aperis, P. M. Oppeneer, and M. V. Miloševi{\'{c}}, Phys. Rev. Lett. {\bf 123}, 077001 (2019).
\bibitem{h-c} G. Savini, A. C. Ferrari and F. Giustino, Phys. Rev. Lett. {\bf 105}, 037002 (2010).
\bibitem{h-b} Y. J. Chen, H. Y. Lu, F. L. Shao and P. Zhang, Phys. Rev. Mater. {\bf 7}, 034004 (2023).
\bibitem{c1} Z. H. Zhang, X. F. Liu, B. I. Yakobson, W. L. Guo, J. Am. Chem. Soc. \textbf{134}, 19326 (2012).
\bibitem{c2} C. Xu, L. Wang, Z. Liu, L. Chen, J. Guo, N. Kang, X. L. Ma, H. M. Cheng, W. Ren, Nat. Mater. \textbf{14}, 1135 (2015).
\bibitem{graphene-sc}  C. Si, Z. Liu, W. Duan and F. Liu, Phys. Rev. Lett. {\bf 111}, 196802 (2013).
\bibitem{T-graphene} S. X. Qiao, C. H. Sui, L. Yang, Y. P. Li, Y. X. Sun, N. X. Zhang, J. Q. Bai, N. Jiao and H. Y. Lu, Phys. Chem. Chem. Phys. {\bf 24}, 25767 (2022).
\bibitem{biphenylene1}  Q. Fan, L. Yan, M. Tripp, S. Dimosthenous, S. Kachel, M. Chen, A. Foster, U. Koert, P. Liljeroth and J. M. Gottfried, Science \textbf{372}, 852-856 (2021).
\bibitem{biphenylene2}  P. F. Liu, J. Li, C. Zhang, X. H. Tu, J. Zhang, P. Zhang, B. T. Wang and D. J. Singh, Phys. Rev. B \textbf{104}, 235422 (2021).
\bibitem{lic6} G. Profeta, M. Calandra, and F. Mauri, Nature Phys \textbf{8}, 131-134 (2012).
\bibitem{alc8} H. Y. Lu, Y. Yang, L. Hao, W. S. Wang, L. Geng, M. Zheng, Y. Li, N. Jiao, P. Zhang, and C. S. Ting, Phys. Rev. B {\bf 101}, 214514 (2020).
\bibitem{b1} Y. V. Kaneti, D. P. Benu, X. T. Xu, B. Yuliarto, Y. Yamauchi, and D. Golberg, Chemical Reviews \textbf{122}, 1000 (2022).
\bibitem{b2} T. J. Xu, Y. H. Wang, Z. Z. Xiong, Y. T. Wang, Y. J. Zhou, X. F. Li, Nano-Micro Lett. \textbf{15}, 6 (2023).
\bibitem{b3} Q. C. Li et al., Science \textbf{371}, 1143 (2021).
\bibitem{n1} H. L. Zhuang, A. K. Singh, R. G. Hennig, Phys. Rev. B: Condens. Matter Mater. Phys. \textbf{87}, 165415 (2013).
\bibitem{bn} X. T. Jin, X. W. Yan, and M. Gao, Phys. Rev. B {\bf 101}, 134518 (2020).
\bibitem{bn3} H. D. Liu, Y. P. Li, L. Yang, N. Jiao, M. M. Zheng, H. Y. Lu, and P. Zhang, Phys. Rev. B {\bf 105}, 224501 (2022).
\bibitem{h-bn} Takat B. Rawal, L. H. Chang, H. D. Liu, H. Y. Lu, and C. S. Ting, Phys. Rev. Mater. {\bf 6}, 054003 (2022).
\bibitem{tmd1} S. Manzeli, D. Ovchinnikov, D. Pasquier, O. V. Yazyev, and A. Kis, Nat. Rev. Mater. \textbf{2}, 17033 (2017).
\bibitem{tmd2} L. J. Pi, L. Li, K. L. Liu, Q. F. Zhang, H. Q. Li, and T. Y. Zhai, Adv. Funct. Mater. \textbf{29}, 1904932 (2019).
\bibitem{tmd3} R. Samal, C. S. Rout, Adv. Mater. Interfaces \textbf{7}, 1901682 (2020).
\bibitem{tmd4} B. Zhao, D. Y. Shen, Z. C. Zhang, P. Lu, M. Hossain, J. Li, B. Li, X. D. Duan, Adv. Funct. Mater. \textbf{31}, 2105132 (2021).
\bibitem{tmd5} T. Taniguchi, L. Nurdiwijayanto, R. Z. Ma, T. Sasaki, Appl. Phys. Rev. \textbf{9}, 021313 (2022).
\bibitem{mxene1} M. Naguib, V. N. Mochalin, M. W. Barsoum,Y. Gogotsi, Adv. Mater. \textbf{26}, 992 (2014).
\bibitem{mxene2} B. Anasori, M. R. Lukatskaya, Y. Gogotsi, Nat. Rev. Mater. \textbf{2}, 16098 (2017).
\bibitem{mxene3} X. T. Jiang, A. V. Kuklin, A. Baev, Y. Q. Ge, H. Ågren, H. Zhang, P. N. Prasad, Physics Reports \textbf{848}, 1-58 (2020).
\bibitem{mxene4} A. VahidMohammadi, J. Rosen, Y. Gogotsi, Science \textbf{372}, 1581 (2021). 
\bibitem{tmd6} F. Yang, P. Hu, F. F. Yang, B. Chen, F. Yin, R. Y. Sun, K. Hao, F. Zhu, K. S. Wang, Z. Y. Yin, Adv. Sci. \textbf{10}, 2300952 (2023). 
\bibitem{tmd7} S. Joseph, J. Mohan, S. Lakshmy, S. Thomas, B. Chakraborty, S. Thomas, N. Kalarikkal, Mater. Chem. Phys. \textbf{297}, 127332 (2023).
\bibitem{mos2-experiment} Kin Fai Mak, Changgu Lee, James Hone, Jie Shan, and Tony F. Heinz, Phys. Rev. Lett. {\bf 105}, 136805 (2010).
\bibitem{mos2app1} L. Li, Y. K. Zhang, D. X. Shi, G. Y. Zhang, Acta Phys. Sin. \textbf{71}, 108102 (2022).
\bibitem{mos2app2} R. Ganatra, Q. Zhang, ACS Nano \textbf{8}, 4074–4099 (2014).
\bibitem{mos2app3} X. Li, H. Zhu, J Materiomics \textbf{1}, 33–44 (2015).
\bibitem{mxene5} M. Naguib, M. W. Barsoum, Y. Gogotsi, Adv. Mater. \textbf{33}, 2103393 (2021). 
\bibitem{mxene6} S. Saharan, U. Ghanekar, and S. Meena, ChemistrySelect \textbf{7}, e202203288 (2022).
\bibitem{mxene7} K. Khan, A. K. Tareen, M. Iqbal, I. Hussain, A. Mahmood, U. Khan, M. F. Khan, H. Zhang, Z. J .Xie, J. Mater. Chem. A \textbf{11}, 19764 (2023)
\bibitem{mxene8} L. M. Malaki, X. T. Jiang, H. L. Wang, R. Podila, H. Zhang, P. Samorì, R. S. Varma, Chem. Eng. J. \textbf{463}, 142351 (2023).
\bibitem{mo2c} C. Xu, L. Wang, Z. Liu, L. Chen, J. Guo, N. Kang, X. L. Ma, H. M. Cheng, W. Ren, Nat. Mater. \textbf{14}, 1135 (2015).
\bibitem{yxz} X. Z. Yin, H. Wang, Q. H. Wang, N. Jiao, M. Y. Ni, M. M. Zheng, H. Y. Lu and P. Zhang, Chinese Phys. Lett. \textbf{40}, 097404 (2023).
\bibitem{wanghao} H. Wang, X. Z. Yin, L. Yang, Y. P. Li, M. Y. Ni, N. Jiao, H. Y. Lu and  P. Zhang, Phys. Chem. Chem. Phys. \textbf{25}, 22171 (2023).
\bibitem{topology} Y. Liang, M. Khazaei, A. Ranjbar, M. Arai, S. Yunoki, Y. Kawazoe, H. Weng, Z. Fang, Phys. Rev. B \textbf{96}, 195414 (2017).
\bibitem{mag1} S. Chen, Z. Jian and Z. Sun, ACS Appl. Mater. Interfaces \textbf{7}, 17510 (2015).
\bibitem{mag2} G. Gao, G. Ding, J. Li, K. Yao, M. Wu and M. Qian, Nanoscale \textbf{8}, 8986 (2016).
\bibitem{sm} See Supplemental Material for calculation details, structural parameters, cohesive energy, AIMD, phonon spectra, band, DOS, band topology, superconductivity, CDW and magnetism for stable MSenes.
\bibitem{hf2s-exp} Kang et al., Sci. Adv. \textbf{6}, eaba7416 (2020).
\bibitem{cu2si1} L. M. Yang, V. Bacic, I. A. Popov, A. I. Boldyrev, T. Heine, T. Frauenheim and E. Ganz, J. Am. Chem. Soc. \textbf{137}, 2757 (2015).
\bibitem{cu2si2} B. Feng, et al., Nat. Commun. \textbf{8}, 1007 (2017).
\bibitem{gas} S. Demirci, N. Avazli, E. Durgun, and S. Cahangirov, Phys. Rev. B {\bf 95}, 115409 (2017).
\bibitem{hf2s-cal} J. F. Zhang, D. Xu, X. L. Qiu, N. N. Zhao, Z. Y. Lu, and K. Liu, J. Phys. Chem. C \textbf{127}, 696 (2023).
\end{thebibliography}
\end{document}